\UseRawInputEncoding

\documentclass[letterpaper, 10 pt, conference]{ieeeconf}  

\IEEEoverridecommandlockouts                              

\usepackage{graphicx} 
\usepackage{amsmath} 
\usepackage{amssymb}  

\usepackage{siunitx}
\usepackage{subfigure}
\usepackage{floatrow}
\usepackage[absolute]{textpos}
\title{\LARGE \bf
Toward Unsupervised Test Scenario Extraction for Automated Driving Systems from Urban Naturalistic
Road Traffic Data}

\author{Nico Weber$^{1,2}$, Christoph Thiem$^{1}$, and Ulrich Konigorski$^{2}$
\thanks{$^{1}$ Opel Automobile GmbH, Stellantis NV, 65423 R\"usselsheim am Main, Germany
        {\tt\small nico.weber@external.stellantis.com}}%
\thanks{$^{2}$ Control Systems and Mechatronics Laboratory, Technical University of Darmstadt, 64283
	Darmstadt, Germany}%
}

\usepackage[pscoord]{eso-pic}
\newcommand{\placetextbox}[3]{
	\setbox0=\hbox{#3}
	\AddToShipoutPictureFG*{ \put(\LenToUnit{#1\paperwidth},\LenToUnit{#2\paperheight}){\vtop{{\null}\makebox[0pt][c]{#3}}}
	}
}
\placetextbox{.42}{0.055}{\small{Preprint of accepted version submitted to SAE International Journal of Connected and Automated Vehicles}}

\begin{document}


\maketitle
\thispagestyle{empty}
\pagestyle{empty}

\begin{abstract}

Scenario-based testing is a promising approach to solve the challenge of proving the safe behavior of vehicles equipped with automated driving systems. Since an infinite number of concrete scenarios can theoretically occur in real-world road traffic, the extraction of scenarios relevant in terms of the safety-related behavior of these systems is a key aspect for their successful verification and validation. Therefore, a method for extracting multimodal urban traffic scenarios from naturalistic road traffic data in an unsupervised manner, minimizing the amount of (potentially biased) prior expert knowledge, is proposed. Rather than an (elaborate) rule-based assignment by extracting concrete scenarios into predefined functional scenarios, the presented method deploys an unsupervised machine learning pipeline. The approach allows exploring the unknown nature of the data and their interpretation as test scenarios that experts could not have anticipated. The method is evaluated for naturalistic road traffic data at urban intersections from the inD and the Silicon Valley Intersections datasets. For this purpose, it is analyzed with which clustering approach (\textit{K}-Means, hierarchical clustering, and DBSCAN) the scenario extraction method performs best (referring to an elaborate rule-based implementation). Subsequently, using hierarchical clustering the results show both a jump in overall accuracy of around 20\% when moving from 4 to 5 clusters and a saturation effect starting at 41 clusters with an overall accuracy of 84\%. These observations can be a valuable contribution in the context of the trade-off between the number of functional scenarios (i.e., clustering accuracy) and testing effort. Possible reasons for the observed accuracy variations of different clusters, each with a fixed total number of given clusters, are discussed. The findings encourage the use of this type of data and unsupervised machine learning approaches as valuable pillars for a systematic construction of a relevant scenario database with sufficient coverage for testing automated driving systems.

\end{abstract}

\section{INTRODUCTION}

The maturity of technical implementations of automated driving systems (ADS)~\cite{SocietyofAutomotiveEngineers.} and resulting fields of application are continuously increasing. While ADS-equipped vehicles promise to contribute significantly to a safer, more efficient, and more comfortable future mobility~\cite{Wood.2019}, the greatest challenge for a market launch of such systems arises from the need for prior proof of safety of the intended functionality~\cite{InternationalOrganizationforStandardization.30.06.2022} for future operation in real-world road traffic~\cite{Hallerbach.2018, Wachenfeld.2016}. Regarding to an urban operational design domain (ODD), possible increases in traffic safety are particularly relevant, as nearly 70\% of accidents involving personal injury in Germany do occur in urban areas~\cite{StatistischesBundesamt.2021} (data collection period: 2016-2020).

Since safety validation approaches currently applied to driving automation systems (SAE Level $<$ 3~\cite{SocietyofAutomotiveEngineers.}) would require billions of representative test kilometers before market launch~\cite{Wachenfeld.2016}, new approaches to safety validation of ADS-equipped vehicles are under development. One of these approaches is the scenario-based development and test approach, as proposed inter alia by project PEGASUS~\cite{Mazzega.2019}. There are several other projects and efforts related to the scenario-based approach such as DARPA Assured Autonomy (US)~\cite{DefenseAdvancedResearchProjectsAgency.}, Scenic initiated at UC Berkeley (US)~\cite{Fremont.2019}, SAKURA (Japan)~\cite{SAKURAProjectConsortium.}, Zenzic (UK)~\cite{ZenzicProjectConsortium.}, and MOOVE (France)~\cite{MOOVEProjectConsortium.}. Following the paradigm of this approach, a high amount of the test effort is shifted to simulation-based approaches~\cite{Pilz.2019,Nalic.2021,Hallerbach.2022} ranging from model-in-the-loop over various X-in-the-loop test methods (e.g., hardware-in-the-loop, prototype-in-the-loop)~\cite{Hallerbach.2018,Steimle.2022} to real-world driving on proving grounds and public roads, covering representative and relevant scenarios~\cite{Wood.2019,Ulbrich.2015}. In recent years, research into safety validation for ADS has primarily been focused on highway applications (e.g., a Highway Chauffeur)~\cite{Mazzega.2019}. With the approval of the first Level 3 ADS for traffic jam situations within highway environments by German Federal Motor Transport Authority (KBA) by end of 2021~\cite{DaimlerAG.2021}, series-production ADS-equipped vehicles are expected to be introduced on public roads soon.

However, both the definition of "relevant scenarios" and a commonly accepted methodology for deriving a scenario database of sufficient coverage are subject of ongoing research~\cite{Wood.2019,Riedmaier.2020}. Existing work on deriving scenarios for the creation of such a scenario database can be generally divided into approaches explicitly generating new scenarios (e.g., leveraging traffic simulations~\cite{Weber.2021} or adversarial testing~\cite{Feng.2021,RamakrishnaShreyas.2022,Eberle.31.10.2019}) or approaches extracting scenarios from recorded data (e.g., accident data~\cite{Sander.2018} or naturalistic road traffic data~\cite{Ries.2021,King.2021}), each of which can be knowledge-based, data-driven or a combination of both~\cite{Riedmaier.2020,DeminNalic.}. The method presented in this paper is to be assigned to data-driven scenario extraction. The basic challenge, independent of the ODD, is the enormously high (in principle infinite) number of traffic situations that the ADS must cope with, referred to as open-world problem~\cite{Stellet.2019}. The goal of a scenario extraction method can be summarized as mapping of this infinite-dimensional open world to a finite and manageable set of scenarios~\cite{Neurohr.2021} that reflects the nature of the traffic dynamics of interest in a sufficiently valid manner for subsequent testing.

Real-world data represents a valuable data source for constructing a comprehensive scenario database~\cite{Wood.2019,Putz.2017}. While the construction of such a database solely by expert knowledge already seems extremely challenging for highway environments, even when applying sophisticated statistical approaches~\cite{Weber.2020}, it appears to be virtually impossible for urban ODDs. This additional increase in complexity when transitioning to urban ODDs is due to significant changes regarding the PEGASUS six-layer model for structuring scenarios~\cite{Scholtes.2021} for the first four layers. The main reason for these changes is a substantially increasing variability in terms of both traffic spaces and traffic dynamics. Particularly, the less rule-based behavior and multimodal interaction of various road user types are to be considered as crucial aspects.

Existing publications concerning data-driven scenario extraction can be divided into rule-based approaches~\cite{Weber.2021,Hartjen.2019}, machine learning based approaches~\cite{Kruber.2019,Watanabe.2019}, or a combination of both~\cite{Ponn.2020,PhilipElspas..2021}. Rule-based approaches have the advantage of not relying on large amounts of annotated (labeled) data. However, the complexity of the rules to be implemented and thus the required effort increases with the complexity of the reality of interest~\cite{NationalAeronauticsandSpaceAdministration.July072016} to be captured. On the one hand, this leads to potentially undetected known, unsafe scenarios. On the other hand, it is impossible to explore previously unknown, unsafe scenarios. Supervised machine learning approaches, in turn, require a sufficient amount of labeled data. As these labels are partly unknown in advance, discovering them must be an intrinsic part of a comprehensive scenario extraction method before respective supervised approaches can be deployed. For this purpose, the characteristics of unsupervised machine learning approaches can be leveraged. In contrast to rule-based approaches, the a priori unknown set of functional scenarios for the examined ODD does not have to be defined in advance. Rather, the definition of a set of functional scenarios is derived by interpreting the results of an unsupervised machine learning approach, such as a cluster analysis, itself. Thereby, all concrete scenarios belonging to a cluster can be assigned to a functional scenario and obtain the corresponding label, for example for the subsequent training of a supervised approach. However, developing a powerful and robust unsupervised machine learning approach is challenging since clustering is a subjective process in nature due to the absence of a ground truth~\cite{King.2021,Xu.2008}. The main goal of the clustering and the subsequent result interpretation itself is to find these before unknown labels as good as possible within the given data. Clustering results heavily depend on the choice of the clustering approach, and even for the same algorithm, the selection of related parameters~\cite{Xu.2008}. Furthermore, clustering results are a matter of view, and the definition of similarity highly depends on the problem. A significant advantage of the unsupervised approach is the fact that in addition to known, unsafe scenarios, previously unknown, unsafe scenarios can be revealed, which are of great importance for a reliable safety argumentation.

\begin{figure*}[t]
	\centering
	\includegraphics[width=4.5in]{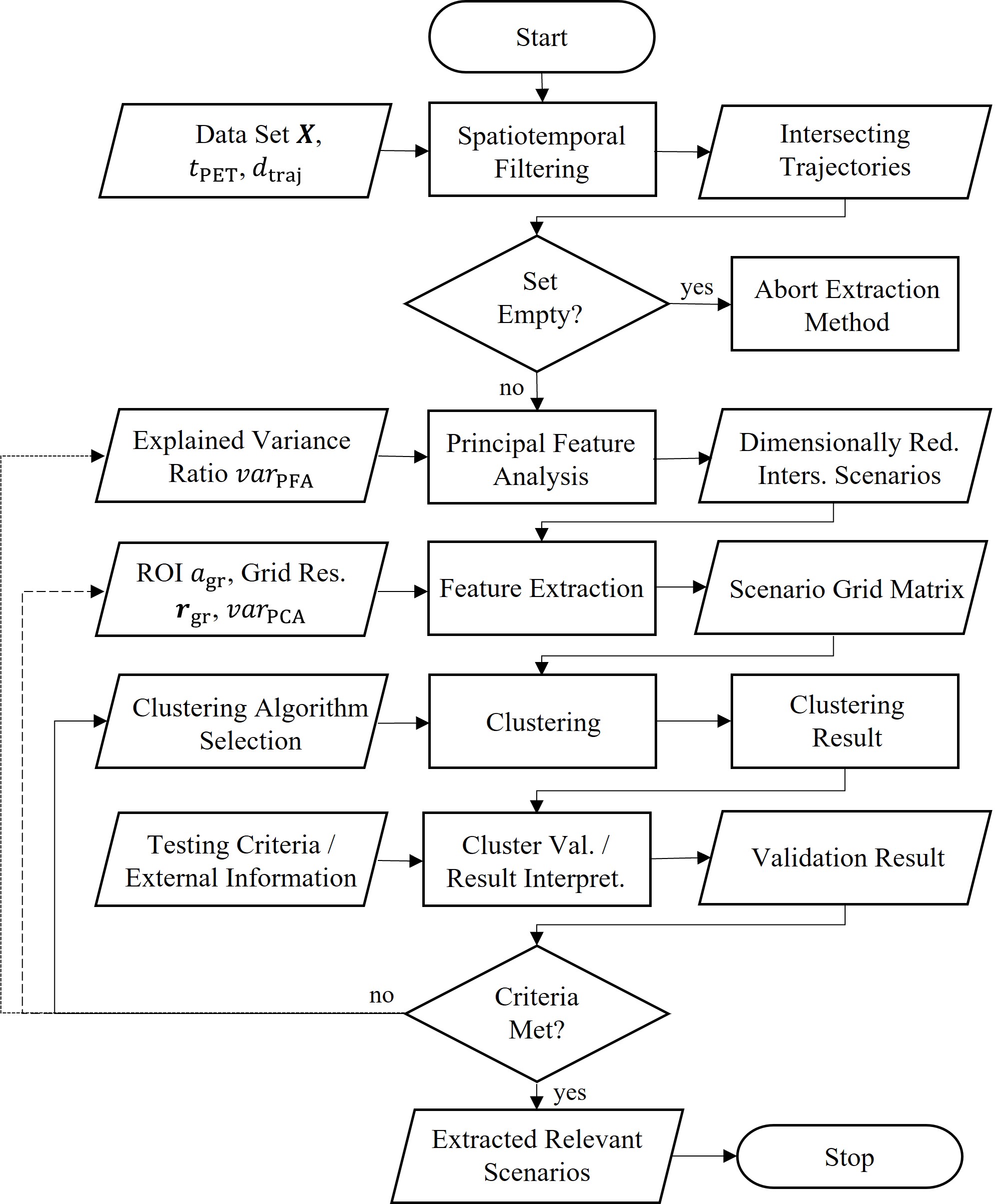}
	\caption{Method for extracting multimodal urban traffic scenarios (dashed lines: optional).}
	\label{fig_method}
\end{figure*}

Therefore, this paper proposes a method for extracting multimodal urban traffic scenarios from naturalistic road traffic data in an unsupervised manner, minimizing the amount of prior expert knowledge required, as shown in Figure~\ref{fig_method}. Based on knowledge gained from rule-based investigations~\cite{Weber.2021}, we build on the generic clustering procedure according to Xu and Wunsch~\cite{Xu.2008}. The method contains novel implementations for all generic sub-steps of a cluster analysis to allow for exploring the unknown nature of the data~\cite{Everitt.2011} and their interpretation as multimodal urban traffic scenarios relevant for testing automated driving systems. The first step of the method involves spatiotemporal filtering of the given dataset, which reduces noise in the data by removing irrelevant data content (e.g., vehicles driving straight through the observation area without other road users present). Afterwards, principal feature analysis (PFA)~\cite{Lu.2007} is applied to the filtered dataset. This step serves the purpose of selecting the most relevant features for the subsequent process steps. The following feature extraction proposes the use of scenario grids for a scenario representation compatible with the application of static data based clustering algorithms and deploys principal component analysis (PCA) on this scenario representation. We are emphasizing a scenario representation compatible with established approaches in the field of computer vision, like convolutional neural networks, for a future deployment of the method within a semi-supervised machine learning pipeline. Based on this feature extraction approach the method allows for comparing different clustering approaches to select the most powerful one for the given dataset. To apply the method exemplarily, hierarchical agglomerative clustering is utilized. For the validation of the clustering results and the result interpretation, the method proposes the comparison with a rule-based reference and introduces so-called cluster validation plots. The step of result interpretation using the cluster validation plots finally leads to extracted relevant traffic scenarios for testing ADS including multimodal interaction of various road users. The method is evaluated exemplarily using the inD dataset~\cite{Bock.2020} and the Silicon Valley Intersections dataset~\cite{levelXdata.2020} and shows promising results. It appears that unsupervised machine learning can make a valuable contribution to constructing a relevant scenario database with sufficient coverage for testing ADS.

This paper is structured as follows. In Section~\ref{Background}, we provide an overview of existing time series clustering approaches and data-driven scenario extraction approaches. Subsequently, in Section~\ref{Method}, we introduce the novel method for unsupervised relevant scenario extraction from naturalistic road traffic data for testing ADS. In Section~\ref{Eval}, the exemplary implementation and evaluation of the scenario extraction method is described. This is followed in Section~\ref{discussion} by a discussion of the results with regard to the fulfillment of the stated requirements through the proposed method. Finally, Section~\ref{conclusion} entails an elaboration on the results showing future research directions.

\section{Background} \label{Background}

\subsection{Requirements Specification}
The specification of requirements for the scenario extraction method should be accompanied by a prioritization of them to develop a common understanding and to achieve a focus on the most important aspects of the development task at hand. We utilize the MoSCoW prioritization, where each requirement is marked as must (M), should (S), could (C), or won´t (W) according to its importance. Note that requirements marked as W are potentially as important as the ones marked as M, but they can be left for a future release of the implementation~\cite{Khan.2015}. The following requirements for the scenario extraction method have been identified by involving OEM domain experts, domain experts from academia as well as relevant standards~\cite{InternationalOrganizationforStandardization.30.06.2022} and regulations~\cite{UnitedNationsEconomicComissionforEurope.09.03.2021,EuropeanCommission.5.8.2022}:
\begin{itemize}
\item The scenario extraction method must allow the processing of naturalistic road traffic data by means of (multivariate) time series of different lengths (M1).
\item The scenario extraction method must allow the clustering of multimodal urban traffic scenarios of a varying number of road users (M2).
\item The scenario extraction method must be applicable to different urban traffic spaces with as little adaptation effort as possible (M3).
\item The scenario extraction method must include a feature representation that is compatible with supervised machine learning approaches (M4).
\item The scenario extraction method must allow a slicing of time series data in meaningful sub-sequences (M5).
\item The scenario extraction method should have a feature representation that includes multiple road user dynamics simultaneously at the scenario level (S1).
\item The scenario extraction method should minimize the amount of prior expert knowledge required for its application (S2).
\item The scenario extraction method is intended to enable scenario clustering without including road network information (C1).
\item The scenario extraction method is intended to enable scenario clustering based on different data sources in future releases (W1).
\end{itemize}

\subsection{Clustering of Time Series Data}

The major strategies for clustering time series data can be divided into raw-data based, feature-based, and model-based approaches~\cite{Liao.2005}. Raw-data based (direct) approaches typically compare different time series using established clustering algorithms by replacing the distance or similarity measure for static data with an appropriate one for time series, e.g., dynamic time warping~\cite{Xu.2008}.

Model-based approaches assume that the time series under investigation are based on an underlying model or that they are determined by a combination of probability distributions and are evaluated by means of a similarity measure between fitted models~\cite{Maharaj.2019}. Both model-based and feature-based approaches can be categorized as indirect approaches, as they first convert raw time series data into a feature vector of lower dimension and perform the clustering on the model parameters or the feature vector, respectively, rather than on the raw time series data themselves~\cite{Liao.2005}. Feature-based (representation-based~\cite{Rani.2012}) approaches can be further divided into strategies deploying manual or automated feature extraction (Representation Learning~\cite{WangWenshuo.2020}).

Despite the enormous number of existing approaches within the major strategies described, we are not aware of any implementation that meets all our requirements specified. This fact is mainly caused by three characteristics within the nature of our problem at hand: First, the scenario extraction method must be able to deal with time series of different lengths, as different road users have, e.g., different velocities while passing through the observation area (cf. Req. M1). Second, the method must cope with multivariate time series since a relevant scenario is composed of at least two road users potentially involving multiple time series per road user for a meaningful scenario representation (cf. Req. M2). Third, the extraction method must cope with a varying amount of time series to be considered because semantically similar scenarios can include different numbers of road users (cf. Req. M2 and S1). Therefore, we deploy a novel approach for scenario representation that is tailored to the problem at hand, allowing the use of established clustering algorithms for static data.

\subsection{Related Work -- Data-Driven Scenario Extraction}

Most of the approaches dealing with data-driven scenario extraction are developed for highway use cases (e.g., \cite{Hallerbach.2018,Kruber.2019,Watanabe.2019,Gelder.2017,Erdogan.2019}). Elpsas \textit{et al.}~\cite{PhilipElspas..2021} propose a method where rule-based detections are used to train fully convolutional networks for extracting lane change and cut-in scenarios. Kerber \textit{et al.}~\cite{Kerber.2020} define a custom scenario distance measure, considering the occupancy and relative distances of vehicles in eight slots surrounding the ego-vehicle, for subsequent hierarchical agglomerative clustering. The authors mention that the method is not suited for urban scenario extraction without modifications. Montanari \textit{et al.}~\cite{Montanari.2020} encourage the approach of extracting scenarios from real driving data in an unsupervised manner. They describe a method for slicing bus communication data of test vehicles, extract features from these slices and cluster them into different highway scenarios using hierarchical agglomerative clustering.

As the traffic dynamics within urban ODDs are characterized by multimodal interactions of various road user types, the scenario extraction method presented in this paper shall be particularly able to extract scenarios based on naturalistic road traffic data, e.g., recorded by unmanned aerial vehicles (drones). This type of data collection allows an uninfluenced and objective view of a scene~\cite{Ulbrich.2015} and thus enables capturing the traffic dynamics that an ADS-equipped vehicle is exposed to in real-world road traffic~\cite{Bock.2020}. Publications dealing with an urban ODD mostly focus on scenario extraction based on other data sources, e.g., accident data or data of field tests with measurement vehicles, and particularly analyze vehicle-to-vehicle interaction scenarios (e.g.,~\cite{Hartjen.2019,WangWenshuo.2020,Barbier.2017}).

Regarding valuable work in terms of scenario extraction based on naturalistic road traffic data, King \textit{et al.}~\cite{King.2021} propose an approach for deriving logical vehicle-to-vehicle interaction scenarios for an unsignalized intersection. Finally, Ries \textit{et al.}~\cite{Ries.2021} introduce a raw-data based clustering method for grouping real driving sequences into semantically similar sequences. The proposed method is the only one we are aware of that leverages clustering to extract urban traffic scenarios with different road user types and numbers, including potentially differing sequence lengths. While these properties lead to the fulfillment of several of the stated must and should requirements (cf. Req. M1, M2, and S2), others of them are not (fully) addressed. In detail, Ries \textit{et al.} state that the method must be adapted for an inter-traffic-space clustering and describe possible solutions to this challenge (cf. Req. M3). Furthermore, we are targeting a feature representation that is compatible with supervised machine learning approaches (cf. Req. M4) and include a slicing of the time series data since one ego-vehicle can encounter multiple concrete scenarios when passing through the observation area (cf. Req. M5)~\cite{King.2021}. Furthermore, instead of a sequential and pairwise similarity estimation between trajectories belonging to a concrete scenario, we want to compare the dynamics of multiple road users simultaneously (cf. Req. S1). Summarizing, to the best of our knowledge, there is no commonly accepted method for extracting relevant multimodal urban traffic scenarios for testing ADS-equipped vehicles that meets all the requirements specified.

\section{Scenario Extraction Method}\label{Method}

The method for extracting multimodal urban traffic scenarios developed with respect to the specified requirements (cf.~Sec.~\ref{Background}) is shown in Figure~\ref{fig_method}. We build on the generic clustering procedure according to Xu and Wunsch~\cite{Xu.2008}. The method contains novel implementations for all generic sub-steps of a cluster analysis to meet the goal of extracting multimodal urban traffic scenarios relevant for testing automated driving systems. For feature selection, the method proposes principal feature analysis~\cite{Lu.2007} and applies a new approach to feature extraction with so-called scenario grids. The method allows for comparing different clustering approaches to select the most powerful one for the dataset at hand. For the validation of the clustering results and the result interpretation, the method proposes the comparison with a rule-based reference and introduces so-called cluster validation plots. While the amount of data decreases with the passing of the different steps, the knowledge of the inherent patterns regarding the reality of interest (i.e., relevant scenarios concerning specified urban traffic spaces) increases. Since the method in the first stage of development must be able to process naturalistic road traffic data particularly, exemplary explanations of the application refer to this type of data source.

\subsection{Spatiotemporal Filtering}


Naturalistic road traffic data contain partly irrelevant parts for subsequent scenario extraction (e.g., vehicles driving straight through the observation area without other road users present). Given a non-empty set of input patterns in form of multivariate time series data~$\textbf{\textit{X}}=\{\textbf{\textit{x}}_{1},...,\textbf{\textit{x}}_{j},...,\textbf{\textit{x}}_{N}\}$, where~$\textbf{\textit{x}}_{j}=\{x_{j\text{1}},x_{j\text{2}},..,x_{jd}\}~\epsilon~\mathbb{R}^{d}$, with a $d$-dimensional feature space, the goal of this process step is to determine the relevant subset of samples $j$ out of the total number of samples $N$, where $j<N$. Hence, this process step aims at reducing noise in the data by removing the irrelevant data proportions. Within the proposed method, this is achieved by a search algorithm based on the post-encroachment time (PET)~\cite{Cooper.1984} of all possible combinations of an ego-vehicle (ego) trajectory and those of encountered road users (challengers) while passing through the observation area (ego-lifetime). As an ego can encounter multiple concrete scenarios within its lifetime (cf. Req. M5), it must be accounted for a slicing of the corresponding trajectories in such cases, additionally.

In detail, this step is performed by a two-layered procedure for each car included in the dataset, as our goal is to extract relevant scenarios for testing ADS-equipped vehicles. In the first layer, a distance matrix is generated for each combination of ego and challenger, containing the Euclidean distances (ego center to challenger center) of the involved trajectories for all common time steps within the observation area. By user-adaptive parametrization of the threshold for the distance between trajectories~$d_{\text{traj}}$, it is possible to define how close the involved trajectories must come to each other so that an interaction between the involved road users is conceivable. The value of~$d_{\text{traj}}$  can be interpreted as radius of a moving circle surrounding the ego, whose area must overlap the challenger trajectory. Within the second layer, the minimum value of the distance matrix values that satisfy this criterion is used to calculate the PET of the trajectories involved. By setting a PET threshold~$t_{\text{PET}}$, the temporal evolvement of the scenario is considered. If the two layers of the algorithm are passed through several times during the lifetime of an ego-vehicle, the resulting ego-challenger combinations can be used to slice the respective time series.

It should be noted that, at the current state of implementation, this algorithm is accompanied by two main assumptions: First, an intersection area (rather than an intersection point) between trajectories involved in a scenario is considered sufficient to maintain it as potentially relevant. Second, the PET between an ego-vehicle and the nearest challenger is crucial for the following classification of the concrete scenario to be kept. In case of an empty set after spatiotemporal filtering, the extraction method is aborted, and, e.g., new data containing relevant scenarios must be recorded for successful application. For further discussion, the entirety of the remaining dataset is referred to as intersecting trajectories.

\subsection{Principal Feature Analysis (Feature Selection)}

The spatiotemporal filtering involves a reduction in the total number of samples $N$, aiming at retaining only relevant traffic scenarios in the dataset. The principal feature analysis aims at reducing the number of columns of the dataset, i.e., reducing the dimension of the feature space $d$. Thus, the task can be described as choosing $n$ distinguishing features out of $d$, whereas $n<d$. An elegant selection of salient features can greatly decrease the storage requirements, simplify the subsequent design process, and facilitate the understanding of the data \cite{Xu.2008,Everitt.2011}.

With respect to the problem at hand, an approach must be chosen that can be used to determine the importance of the features of the intersecting trajectories (e.g., position, orientation, longitudinal and lateral velocity) to subsequently select a meaningful subset of them. It should be noted that we assume that the superset of this meaningful feature subset is contained in the corresponding dataset, since the proposed method is conceptualized for processing naturalistic road traffic data~\cite{Bock.2020}. While there are various methods to reduce the dimensionality of a feature set, e.g., principal component analysis, these approaches are characterized by resulting in a lower-dimensional representation, where the features are not physically interpretable anymore~\cite{Xu.2008,Lu.2007}. While this is unproblematic for other tasks, these approaches are not effective in the context of the subsequently proposed feature representation, where physical interpretability is of great importance.

Hence, the proposed method deploys feature selection by principal feature analysis~\cite{Lu.2007}, which allows to exploit the structure of the principal components of a feature set to find a subset of the original feature vector. The method includes five steps, with the first step calculating the covariance matrix~$\textbf{\textit{C}}$ of the standardized intersecting trajectories dataset~$\textbf{\textit{X}}_{\text{std}}$. This step is followed by calculating the principal components and the eigenvalues of the covariance matrix~$\textbf{\textit{C}}$. By choosing the explained variance ratio~$var_{\text{PFA}}~\epsilon~[0,1]$,~$s$ columns of the matrix~$\textbf{\textit{A}}$, representing the orthonormal eigenvectors of~$\textbf{\textit{C}}$, are kept, constructing the subspace matrix~$\textbf{\textit{A}}_{s}$. The parametrization of~$var_{\text{PFA}}$ decides how much of the variability of the data is desired to be retained. In the third step, the rows~$\textbf{\textit{v}}_{i}~\epsilon~\mathbb{R}^{s}$ of~$\textbf{\textit{A}}_{s}$ are clustered using the \textit{K}-means algorithm. Since each vector~$\textbf{\textit{v}}_{i}$ represents the projection of the $i$\textquotesingle th feature of~$\textbf{\textit{X}}_{\text{std}}$ in the lower-dimensional space, the $s$ elements of~$\textbf{\textit{v}}_{i}$ correspond to the weights of the $i$\textquotesingle th feature on each axis of the subspace $s$. As the amount of mutual information increases with the similarity of the absolute values of these vectors, the clusters derived by \textit{K}-means can be used to choose one feature of a subset of highly correlated features, respectively~\cite{Lu.2007}. According to~\cite{Lu.2007}, these features represent each group optimally by means of high spread in the lower dimension, reconstruction and insensitivity to noise. It is of note that the number of clusters should be chosen slightly higher than $s$.

Finally, applying this method, it is possible to systematically reduce the intersecting trajectories dataset regarding its dimensionality. This paves the way for a well-founded feature extraction, as described below.

\subsection{Feature Extraction}


With the spatiotemporal filtered and dimensionally reduced intersecting trajectories dataset available, the following process step consists of a feature extraction suitable for the subsequent clustering of multimodal urban traffic scenarios. Ideally, such a feature representation should be of use in distinguishing patterns belonging to different clusters, immune to noise, and easy to obtain and interpret~\cite{Xu.2008}. Considering these generic requirements as well as the task-specific requirements (cf. Sec.~\ref{Background}), we propose a two-step process to finally construct a matrix feasible for serving as input for the subsequent process step of clustering.

\begin{figure*}[t]
	\centering
	\includegraphics[width=6.75in]{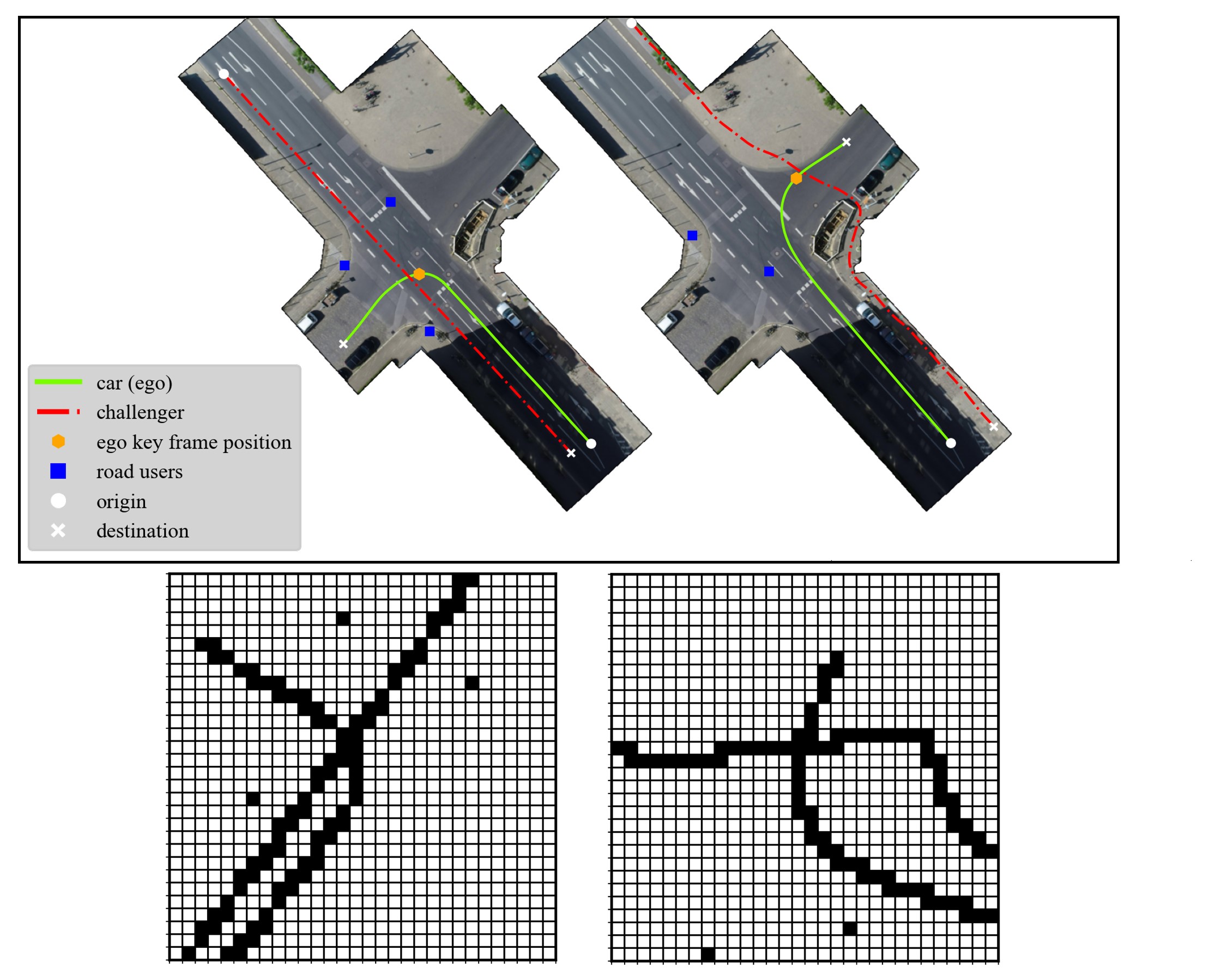}
	\caption{Exemplary concrete scenario visualizations (top) with associated occupancy grid channels (bottom) at Bendplatz traffic space for semantically dissimilar concrete scenarios (left: ego-to-vehicle, right: ego-to-pedestrian).}
	\label{fig_keyframe}
\end{figure*}

We utilize the principle of a grid-based representation, within which a certain spatial area around the ego is discretized and defined as potentially relevant in terms of the ego behavior, as illustrated by Figure~\ref{fig_keyframe} for two semantically dissimilar concrete scenarios at Bendplatz traffic space. The basic principle has already been proposed for use within the robotic domain~\cite{Elfes.1989} and various modules of an automated driving system, such as decision making or motion planning~\cite{Ziegler.2010}. Furthermore, Gruner \textit{et al.}~\cite{Gruner.2017} propose the use of such grid-based representations for scenario classification based on object-list data and confirm the usability of this type of representation for training artificial neural networks. While~\cite{Gruner.2017} suggests the use of a single scene or a fixed number of scenes for scenario classification, we propose the construction of one discrete, multi-channel grid structure per concrete scenario, covering a variable time span and thus number of scenes. The reason for this approach is, on the one hand, the knowledge of the evolvement of an entire scenario based on naturalistic road traffic data. On the other hand, our approach aims at maximizing the automation level of scenario labeling itself by using unsupervised machine learning approaches, while in~\cite{Gruner.2017} a so-called semi-automatic scenario labeling supported by previously defined rules is performed. In addition, the previous spatiotemporal filtering already accounts for the temporal evolvement of a scenario, and scenes are assigned to a respective concrete scenario, even if the label of the corresponding concrete scenario is still unknown. Finally, our investigations showed that the scenario-level representation requires significantly less computational effort.

\subsubsection{Key Frame Calculation}

This step comprises the determination of the point in time or frame, respectively, of the corresponding concrete scenario (key frame), which is used for the subsequent construction of the discrete, multi-channel grid structure, referred to as scenario tensor. On the one hand, the determination of this key frame should be applicable as generically as possible to all relevant ego-challenger combinations identified. On the other hand, the snapshot of the scenario created based on the key frame should capture the distinguishing characteristics of the specific scenario category as accurately as possible.

Our investigations have shown that a computation based on the maximum of the yaw rate of the ego~$\dot{\varphi}_{\text{e,max}}$ or challenger~$\dot{\varphi}_{\text{c,max}}$ within the specific concrete scenario entails the best results. In detail, first, the maximum yaw rates of the road users involved in a concrete scenario are calculated for all ego-challenger combinations within the intersecting scenarios. Then, the maximum of the respective yaw rate value set is calculated. In case of~$\dot{\varphi}_{\text{e,max}}\geq\dot{\varphi}_{\text{c,max}}$, the frame associated with~$\dot{\varphi}_{\text{e,max}}$ is used to construct the scenario tensor. Since the construction of the scenario tensor is always done with respect to the ego-vehicle state, in case of~$\dot{\varphi}_{\text{e,max}}<\dot{\varphi}_{\text{c,max}}$, a shift of time to the associated challenger frame must take place. In the lower part of Figure~\ref{fig_keyframe} the resulting key frames are implicitly shown by the exemplary occupancy grid channels, which originate from two concrete scenarios visualized in the upper part of Figure~\ref{fig_keyframe}.


\subsubsection{Scenario Tensor Construction}

The key frame for each concrete scenario defines which point in time or frame, respectively, is to be used to construct the corresponding scenario tensor~$\Phi=\{\textbf{\textit{G}}_{1},...,\textbf{\textit{G}}_{j},...,\textbf{\textit{G}}_{l}\}$, where~$\textbf{\textit{G}}_{i}$ is the $i$\textquotesingle th grid channel matrix of size~$(a_{\text{gr}} \cdot r_{\text{gr,1}}) \times (a_{\text{gr}} \cdot r_{\text{gr,2}})$. Both the region of interest~$a_{\text{gr}}$ of the rectangular grids and the resolution vector~$\textbf{\textit{r}}_{\text{gr}}=({r_{\text{gr,1}}, r_{\text{gr,2}}})$, containing the longitudinal and lateral grid resolution, respectively, can be adapted to the specific application. While~$a_{\text{gr}}$ determines how far the ego looks into space and time,~$\textbf{\textit{r}}_{\text{gr}}$ determines how fine the grid resolves the spatiotemporal evolvement of the respective scenario. All feature values within the $l$ grid channels are calculated with respect to the ego coordinate system at the key frame. Thus, the scenario tensor comprises a discrete, multi-channel grid structure, representing the spatiotemporal scenario characteristics from a bird´s-eye view. A feasible number of grid channels~$l_{\text{gr}}$ is determined by the result of previous principal feature analysis, with~$l_{\text{gr}}$ less or equal to the dimension of the reduced intersecting trajectories dataset.

In the lower part of Figure~\ref{fig_keyframe}, the occupancy grid~$\textbf{\textit{G}}_{1}$, containing both the trajectories of the ego and the nearest challenger as well as grid cells occupied by other surrounding road users for the corresponding key frame, is visualized for exemplary concrete scenarios. In the context of this work, the nomenclature of the concrete scenarios is based on the combination of ego and nearest challenger, which in case of Figure~\ref{fig_keyframe} leads to the naming ego-to-vehicle (left) as well as ego-to-pedestrian (right) scenario. In agreement with Gruner \textit{et al.}~\cite{Gruner.2017}, this scenario representation is basically applicable to different ODDs and adaptive by means of the region of interest and resolution. Furthermore, other grid forms are conceivable. It should be noted that this type of scenario representation addresses the fulfillment of essential requirements stated (e.g., Reqs. M1, M2, M3, M4, and S1).

To obtain a representation suitable for clustering, each scenario tensor is flattened column-wise into a scenario grid matrix of size~$(a_{\text{gr}} \cdot r_{\text{gr,1}}) \times (a_{\text{gr}} \cdot r_{\text{gr,2}} \cdot l)$. After feature extraction by applying principal component analysis, the entirety of the resulting scenario grids can finally be forwarded for the following process step as cluster input matrix~$\textbf{\textit{M}}_{\text{c}}$.

\subsection{Clustering}


The following process step of the unsupervised scenario extraction method consists of the actual application of a clustering approach. The structure of the cluster input matrix~$\textbf{\textit{M}}_{\text{c}}$ basically allows the application of any clustering algorithm based on static data. Since the scenario extraction method is in general applicable to different data sources and traffic spaces, the user can and must select the most appropriate clustering approach for the dataset at hand.

The evaluation in the scope of this paper shows the most promising clustering results using hierarchical agglomerative clustering~(cf. Sec.~\ref{Eval}). This is in accordance with literature, in which hierarchical agglomerative clustering is described as suitable for use cases with many clusters and many samples~\cite{Lu.2007}. Furthermore, this is in line with other publications in the context of real-data based scenario extraction, both based on similarly structured data sources~\cite{Kerber.2020} as well as for other types of data sources~\cite{Montanari.2020}.

\subsection{Cluster Validation and Result Interpretation}

In literature, three different testing criteria categories, namely external, internal and relative indices, are defined to be able to estimate the quality of clustering results~\cite{Xu.2008}. The first step of our evaluation can be assigned to the relative indices testing criteria category since different clustering approaches and their respective parametrizations are compared against each other. This comparison additionally leverages external indices, where external information is used as standard to validate the clustering results. In detail, the accuracy calculation of each clustering approach relies on a rule-based approach as a reference. In the scope of this paper, the elaborate rule-based implementation specifically designed for Bendplatz traffic space was intentionally accepted to get a better reference to the presented scenario extraction method. The subsequent step of our evaluation can also be assigned to the external indices testing criteria category by utilizing the map information available within the datasets in the form of images of the corresponding traffic spaces. For this purpose, we compare the trajectories involved in different concrete scenarios belonging to a respective cluster by visual validation~(cf. Sec.~\ref{Eval}) in accordance with literature in the field of unsupervised scenario extraction (e.g.,~\cite{Ries.2021,Kerber.2020,Montanari.2020}). Since cluster analysis is not a one-shot process~\cite{Xu.2008}, the method includes feedback loops depending on the validation results for trials and repetitions with an adjusted parametrization at different process steps. In case that the validation criteria are met, the extracted relevant traffic scenarios can be used for subsequent applications, e.g., scenario-based testing of an automated driving system.

\section{Evaluation}\label{Eval}

In this chapter, the method for extracting relevant urban traffic scenarios is evaluated using the inD dataset~\cite{Bock.2020} as well as the Silicon Valley Intersections dataset (sv dataset)~\cite{levelXdata.2020}. Some process steps are evaluated for the entire datasets to get an impression of the generic character of the method. Other process steps are demonstrated for exemplary traffic spaces to illustrate specific advantages and drawbacks of the method in depth.

\subsection{Datasets and Parametrization}

There are various reasons for choosing the inD dataset for the main part of the evaluation of the scenario extraction method. The main advantages of the inD dataset compared to other similar datasets are its size, representativeness, and accuracy~\cite{Bock.2020}. The dataset consists of more than 11,500 naturalistic road user trajectories, including cars, trucks, and busses as well as about 5,000 pedestrian and bicyclist trajectories recorded at four unsignalized intersections in Aachen, Germany. In particular, the high proportion of vulnerable road users makes this dataset an interesting challenger to the method (cf. Tab.~\Ref{ind_overview}), as it can be assumed that it is especially difficult to detect recurring patterns within the less rule-based behavior of vulnerable road users. In addition, we use the Silicon Valley Intersections dataset~\cite{levelXdata.2020} to evaluate individual process steps. This dataset includes naturalistic road user trajectories for seven traffic spaces in the Silicon Valley area. Due to the diversity of the traffic spaces as well as changed boundary conditions (especially concerning layers 1-4 of the six-layer model~\cite{Scholtes.2021}) compared to the inD dataset, more comprehensive conclusions regarding the feasibility of the extraction method can be drawn.

As shown in Tab.~\ref{ind_overview}, Bendplatz traffic space and Frankenburg traffic space of the inD dataset have a high total number of trajectories and a high percentage of vulnerable road users. Hence, these two traffic spaces are particularly interesting for in-depth studies since a higher share of multimodal interaction can be assumed.

\begin{table}[b]
	\caption{InD Dataset Overview}
	\label{ind_overview}
	\begin{center}
		\begin{tabular}{|c||c|c|}
			\hline
			Traffic Space & Share in total & Vulnerable road \\ 
						  & rec. road users & users/Vehicles\\
			\hline
			Heckstra{\ss}e & 0.09 & 0.07\\
			\hline
			Aseag & 0.17 & 0.12\\
			\hline
			Bendplatz & 0.28 & 0.5\\
			\hline
			Frankenburg & 0.46 & 1.6\\
			\hline
		\end{tabular}
	\end{center}
\end{table}

\subsection{Spatiotemporal Filtering and Feature Selection}

For the subsequent studies, the following parametrization was performed:~$t_{\text{PET}}=\SI{3}{\second}$, $d_{\text{traj}}=\SI{1}{\meter}$, $var_{\text{PFA}}=0.95$, $a_{\text{gr}}=\SI{32}{\meter}$, $r_{\text{gr,1}}=r_{\text{gr,2}}=\SI{1}{\text{px}/\meter}$, and $var_{\text{PCA}}=0.99$. It is of note that the parametrization has a high influence on the results obtained and an appropriate choice depends on various factors (e.g., given dataset, examined ODD, etc.). Since the focus is on the proof-of-concept of the method, the numerical values have been exemplarily chosen on the basis of expert judgement and literature references~\cite{Gruner.2017,Patel.2019}.

While the application of the search algorithm for spatiotemporal filtering is trivial, conclusions can already be drawn based on the proportion of the remaining relevant intersecting trajectories dataset. Since the datasets provide a classification regarding the respective road user types (cars, trucks, pedestrians, and bicyclists), high-level scenario categories relevant from the viewpoint of proving the safety of an ADS-equipped vehicle can be defined as ego-to-vehicle (e-to-v) scenarios, ego-to-pedestrian (e-to-p) scenarios, and ego-to-bicyclist (e-to-b) scenarios. For the example of Bendplatz traffic space, on average about 4\% of pedestrians, 14\% of vehicles, and 15\% of bicyclists remain within the dataset for the subsequent process steps. Since a higher proportion of remaining road user types represents a higher proportion of scenarios below the PET threshold~$t_{\text{PET}}$, these numerical values may serve as an indication for the combination of external risk to which the various road user types are exposed and internal road user type-specific risk tolerance. In addition, this gives a hint on the required number of recordings since even with the conservative parametrization performed, most of the original amount of data is classified as not relevant for the exemplary application.

\begin{figure}[t]
	\centering
	\includegraphics[width=3.5in]{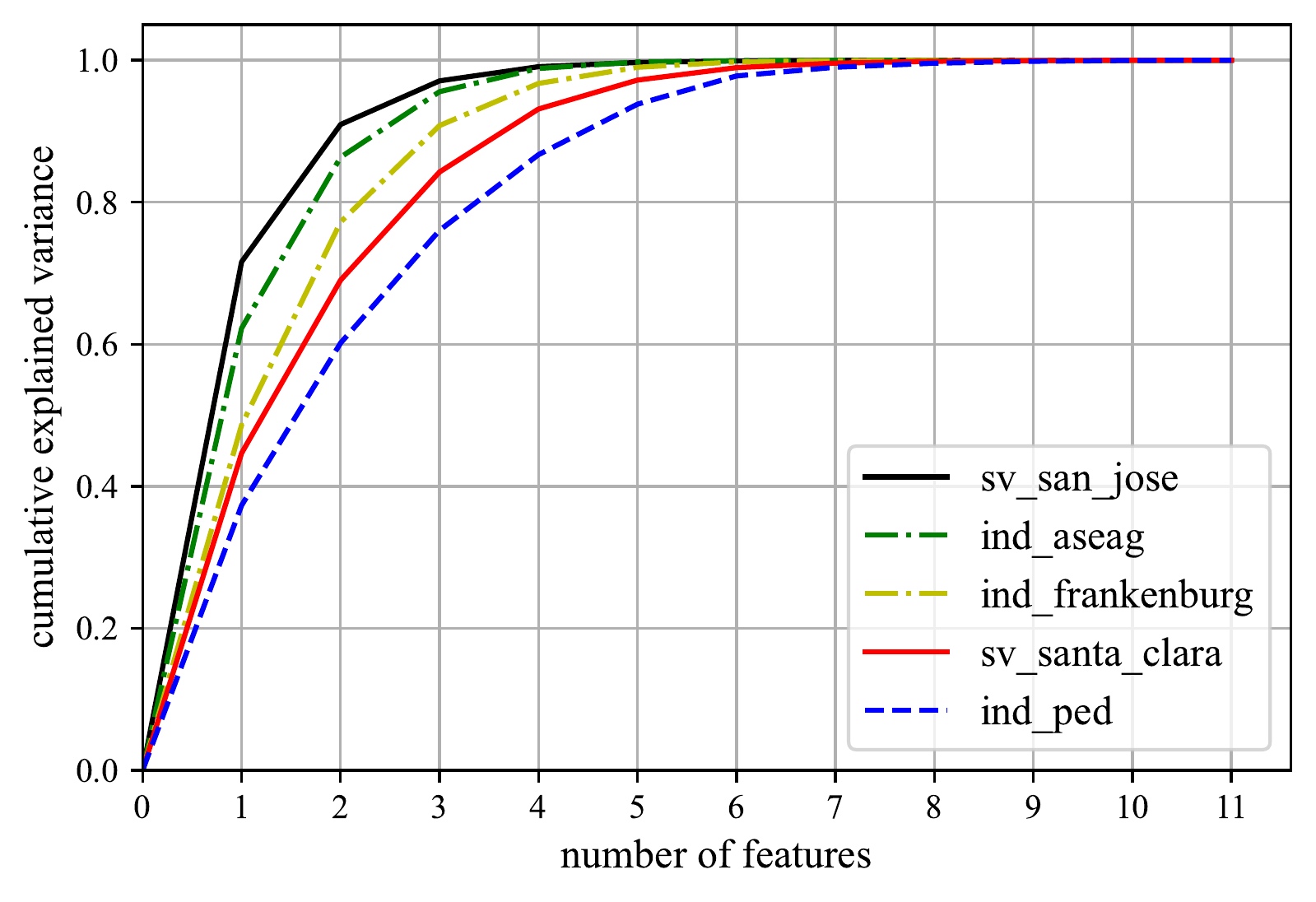}
	\caption{Results of principal feature analysis: cumulative explained variance over the number of selected features for exemplary traffic spaces and road user types.}
	\label{fig_feature_sel}
\end{figure}

Fig.~\ref{fig_feature_sel} shows exemplary results of the principal feature analysis in terms of the cumulative explained variance~$var_{\text{PFA}}$ over the number of features for different traffic spaces and road user types. Note that even if this is not a continuous function, the continuous representation of the graphs supports the understanding of the respective results. Fig.~\ref{fig_feature_sel} shows the curves for the object type car for the traffic spaces with the minimum and maximum number of features required for the defined cumulative explained variance for the corresponding dataset. It is noticeable that the extrema of the sv dataset wrap around those of the inD dataset. This result corresponds to the intuitive expectations since the traffic space \textit{sv\textunderscore san\textunderscore jose} (lat.: 37.3627, long.: -121.8759) is in direct proximity to a highway entrance and only vehicles are present. It is also plausible that traffic space \textit{sv\textunderscore santa\textunderscore clara} (lat.: 37.3252, long.: -121.9490) requires more features for capturing the same cumulative explained variance since it includes a complex parking lot with two 4-way stops and six normal stop entries.

While pedestrians, in general, need more features for the same value of cumulative explained variance, the analysis of the inD dataset shows that the influence of the traffic space is marginal for this road user type, which is why only one curve is shown here (\textit{ind\textunderscore ped}). On average, the eleven features considered (\textit{xCenter}, \textit{yCenter}, \textit{heading}, \textit{xVelocity}, \textit{yVelocity}, \textit{xAcceleration}, \textit{yAcceleration}, \textit{lonVelocity}, \textit{latVelocity}, \textit{lonAcceleration}, \textit{latAcceleration})~\cite{Bock.2020} can be reduced to about the half, with pedestrians defining the lower threshold. This reduction seems intuitively logical for the inD dataset since some features differ only by the reference coordinate system. However, this process step can be an even more valuable tool for objective feature selection for differently structured datasets~\cite{Lu.2007}.

\subsection{Feature Extraction}

\begin{figure*}[t]
	\centering
	\includegraphics[width=6.75in]{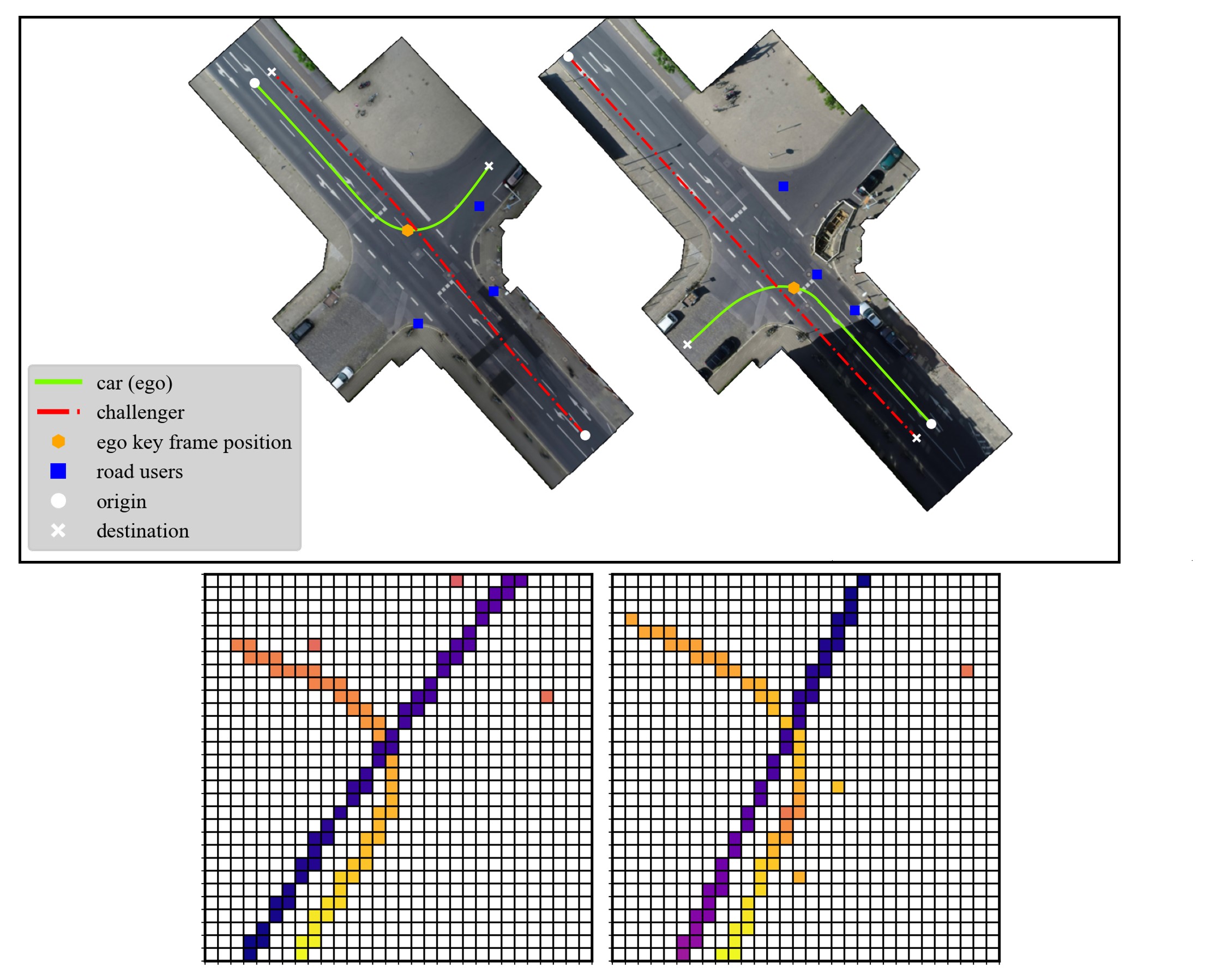}
	\caption{Exemplary concrete scenario visualizations (top) with associated x-velocity grid channels (bottom) for semantically similar ego-to-vehicle scenarios. The concrete scenarios are chosen randomly from one cluster at Bendplatz.}
	\label{fig_grids}
\end{figure*}

The evaluation of the following process steps is conducted exemplarily for Bendplatz traffic space of the inD dataset (cf. Fig.~\ref{fig_keyframe}). The principal feature analysis results in seven features for~$var_{\text{PFA}}=0.95$, with the pedestrian class forming the feature superset of all road user classes. On this basis, the features \textit{xCenter}, \textit{yCenter}, \textit{heading}, \textit{xVelocity}, \textit{yVelocity}, \textit{xAcceleration}, and \textit{yAcceleration} are used to construct the scenario tensor. Since the information of the features \textit{xCenter}, \textit{yCenter}, and \textit{heading} can be combined in the occupancy grid channel, this results in a total of five grid channels consisting of the occupancy grid and velocity and acceleration grids in both spatial directions with respect to the ego state at the respective key frame. While Fig.~\ref{fig_keyframe} shows the deviation of the feature representation for the case of semantically dissimilar scenarios, Fig.~\ref{fig_grids} illustrates the ability of feature representation to represent semantically similar scenarios independent of their absolute position in the traffic space in a similar manner. This is demonstrated in Fig.~\ref{fig_grids} by the \textit{x}-velocity grid channels of two concrete e-to-v scenarios (bottom), which originate from two concrete scenarios visualized in the upper part of Fig.~\ref{fig_grids}. 

\subsection{Cluster Validation and Result Interpretation}


Our approach to validate the clustering results involves three branches (cf. Sec.~\ref{Method}). The first branch consists of a comparison of different clustering approaches and their respective parametrizations, which is to be assigned to the relative indices testing criteria category. Thereby, we present the parametrizations that produced the best results in the scope of this paper. That comparison includes the second branch of the validation by assessing the clustering approaches with respect to an elaborate rule-based implementation, which is to be assigned to the external indices testing criteria category. From this, we aim to gain a broader impression of the performance of the scenario extraction method. Finally, the third branch includes a visual validation approach, which also falls under the external indices testing criteria category.

\subsubsection{Comparison of Clustering Approaches}

\begin{figure}[t]
	\vspace{0.5\baselineskip}
	\centering
	\includegraphics[width=3.4in]{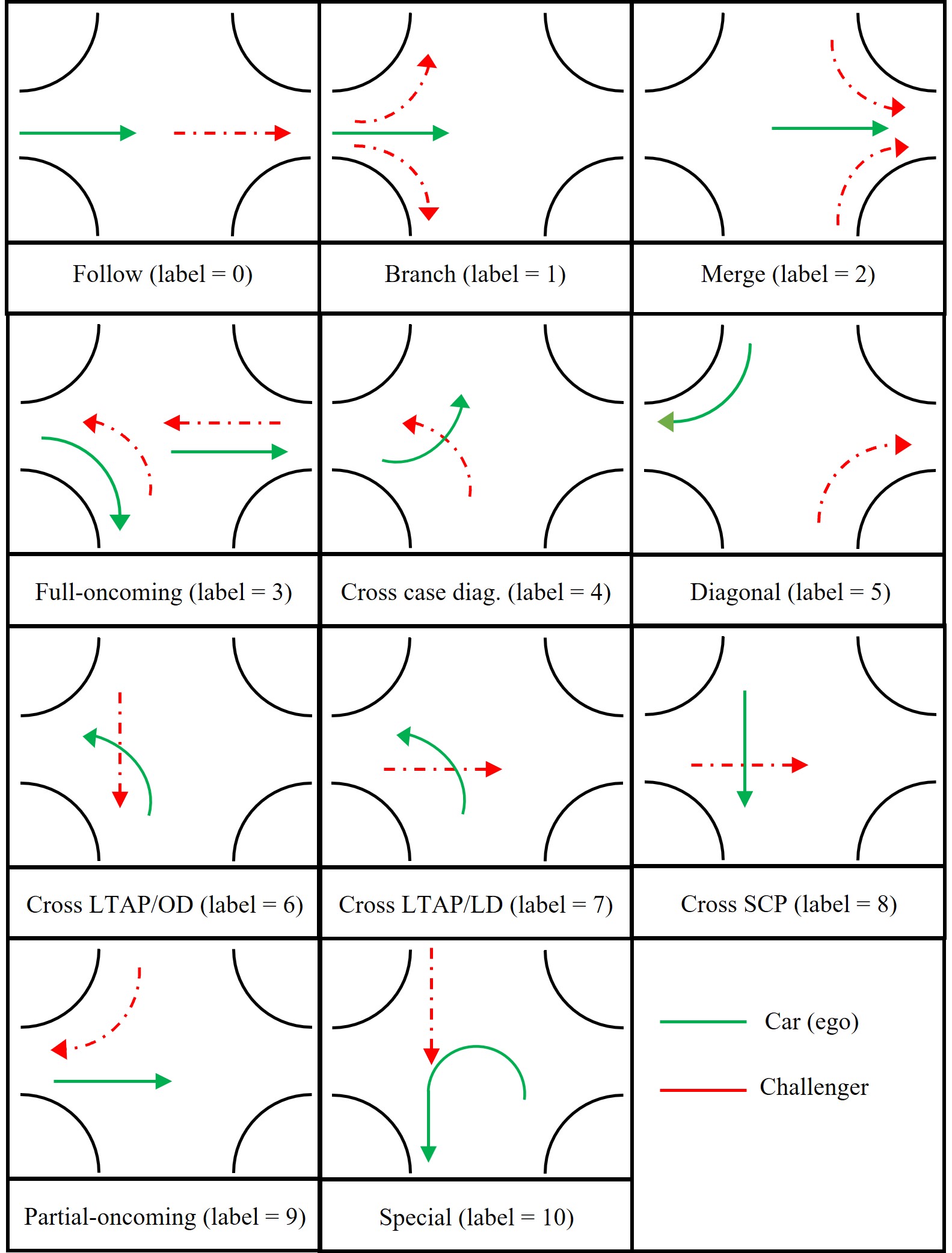}
	\caption{Functional scenario categorization of the rule-based scenario extraction approach. The labels assigned to the samples by this approach serve as reference for the calculation of the overall accuracy and thus for the evaluation of the clustering results based on~\cite{Sander.2018,Cooper.1984}.}
	\label{fig_labels}
\end{figure}

To ensure the selection of a valid clustering approach for the exemplary evaluation of the method, three popular clustering approaches \textit{K}-means, hierarchical agglomerative clustering, and density-based spatial clustering of applications with noise (DBSCAN)~\cite{Xu.2008} are compared in terms of their performance. For this purpose, we compare the results of the clustering process with respect to the e-to-v scenarios for Bendplatz traffic space with a rule-based baseline approach particularly designed for this traffic space~\cite{Weber.2021}. The methodology for determining the overall accuracy is based on Patel~\cite{Patel.2019} and has been extended for this use case. In detail, this involves assigning to each sample one out of the ten associated scenario labels originating from the rule-based approach~\cite{Weber.2021}, which are shown in Figure~\ref{fig_labels}. Outliers are assigned to the "Special" scenario category. An example of such an outlier in the dataset are (illegal) U-turns of vehicles on the corresponding intersection.

\begin{figure*}[t]
	\vspace{0.2\baselineskip}
	\centering
	\includegraphics[width=4.6in]{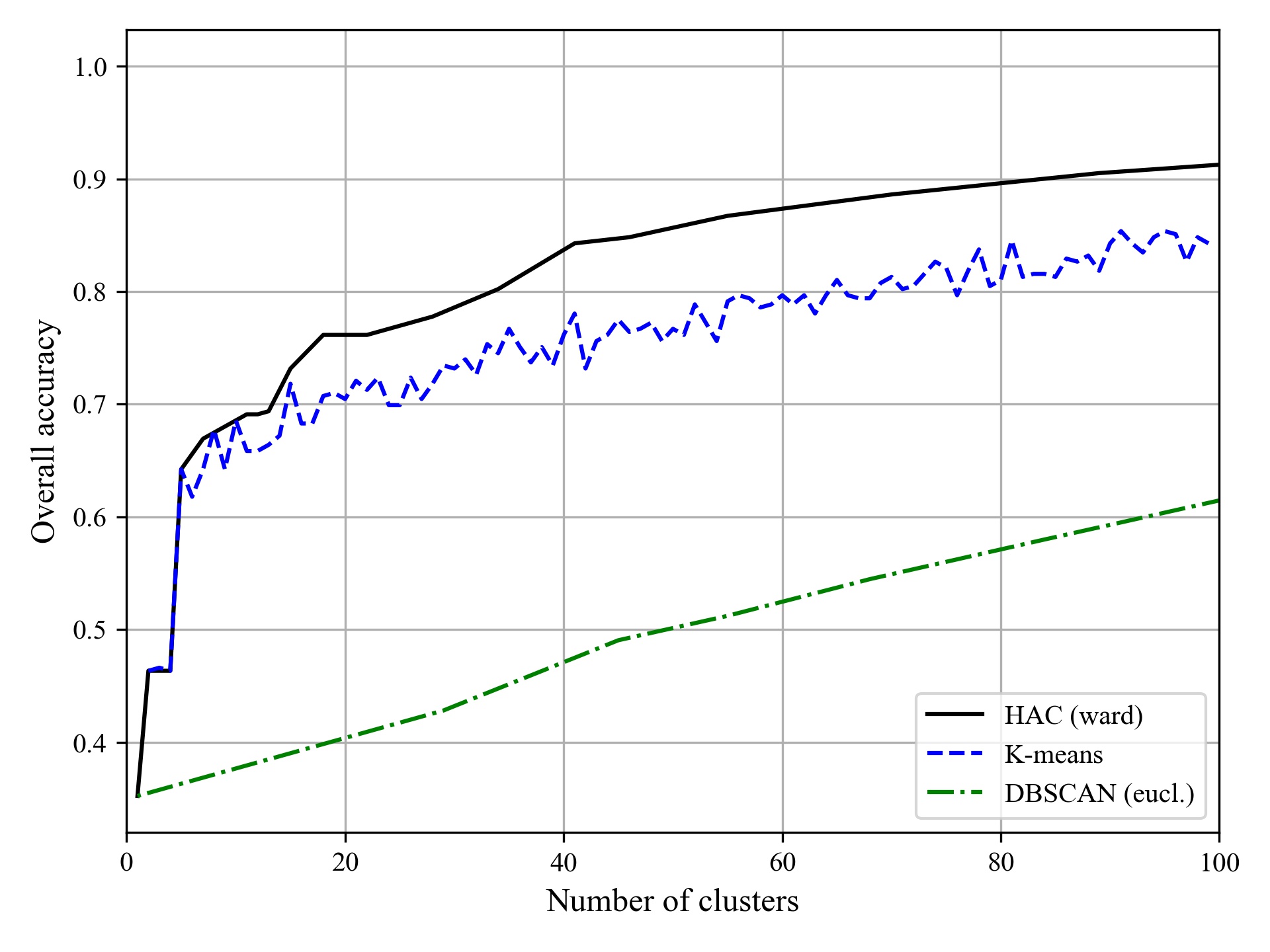}
	\caption{Overall accuracy over the number of clusters for hierarchical agglomerative clustering (HAC), \textit{K}-means clustering, and Density-Based Spatial Clustering of Applications with Noise (DBSCAN) at Bendplatz traffic space.}
	\label{fig_clust_comp}
\end{figure*}

For the subsequent calculation of the overall accuracy, the most frequently occurring labels per cluster are summed up over all clusters and are divided by the total number of samples. This leads to an optimal overall accuracy at a value of 1 or 100\%, respectively. The lower limit of the value range depends on the specific partitioning and the dataset. The worst partitioning into only one cluster leads to an overall accuracy of 0.35 for the e-to-v scenarios at Bendplatz, since this value represents the share of the most frequently occurring scenario category (130 follow scenarios) of the 369 samples available in total. Figure~\ref{fig_clust_comp} shows the resulting overall accuracy for the three clustering approaches over the number of clusters. It becomes clear that hierarchical agglomerative clustering outperforms \textit{K}-means and DBSCAN over the entire spectrum of given clusters. It should be noted that no clustering approach can be generally stated to be best in all circumstances, and performance depends on both the clustering approach and dataset~\cite{Everitt.2011}. In addition, the clustering procedure involves several steps, each of which may depend on the previous one. However, the results are in accordance with the literature~\cite{Xu.2008,Kerber.2020,Montanari.2020} and the use of multiple clustering approaches that provide similar results strengthen the confidence in the results. While hierarchical agglomerative clustering and \textit{K}-means perform similarly depending on the number of clusters, DBSCAN shows significantly worse performance. This is in agreement with Patel~\cite{Patel.2019}, which also compares these three clustering approaches for a similar dataset. For the case of hierarchical agglomerative clustering and \textit{K}-means, there is a jump in the overall accuracy of about 20\% when moving from four to five clusters, which suggests a partitioning into at least five clusters given the goal of defining a database of relevant test scenarios. Regarding the definition of a reasonable upper limit for the number of clusters, the significantly visible saturation of the overall accuracy for hierarchical agglomerative clustering from a cluster number of 41 or an overall accuracy of 84\%, respectively, can be considered. This observation can provide a valuable contribution against the background of the trade-off between clustering accuracy and testing effort. In detail, the clusters, whose number depends on the required overall accuracy, can be used to derive equivalence classes or functional scenarios of a test catalog. The non-determinism of the \textit{K}-means algorithm, due to a variable number of initial centroids with a variable number of clusters, leads to a non-monotonically increasing curve of the corresponding accuracy. Still, the general characteristic of an increasing overall accuracy with an increasing number of clusters is significantly visible.



\begin{figure*}[t]
	\vspace{0.2\baselineskip}
	\centering
	\includegraphics[width=4.6in]{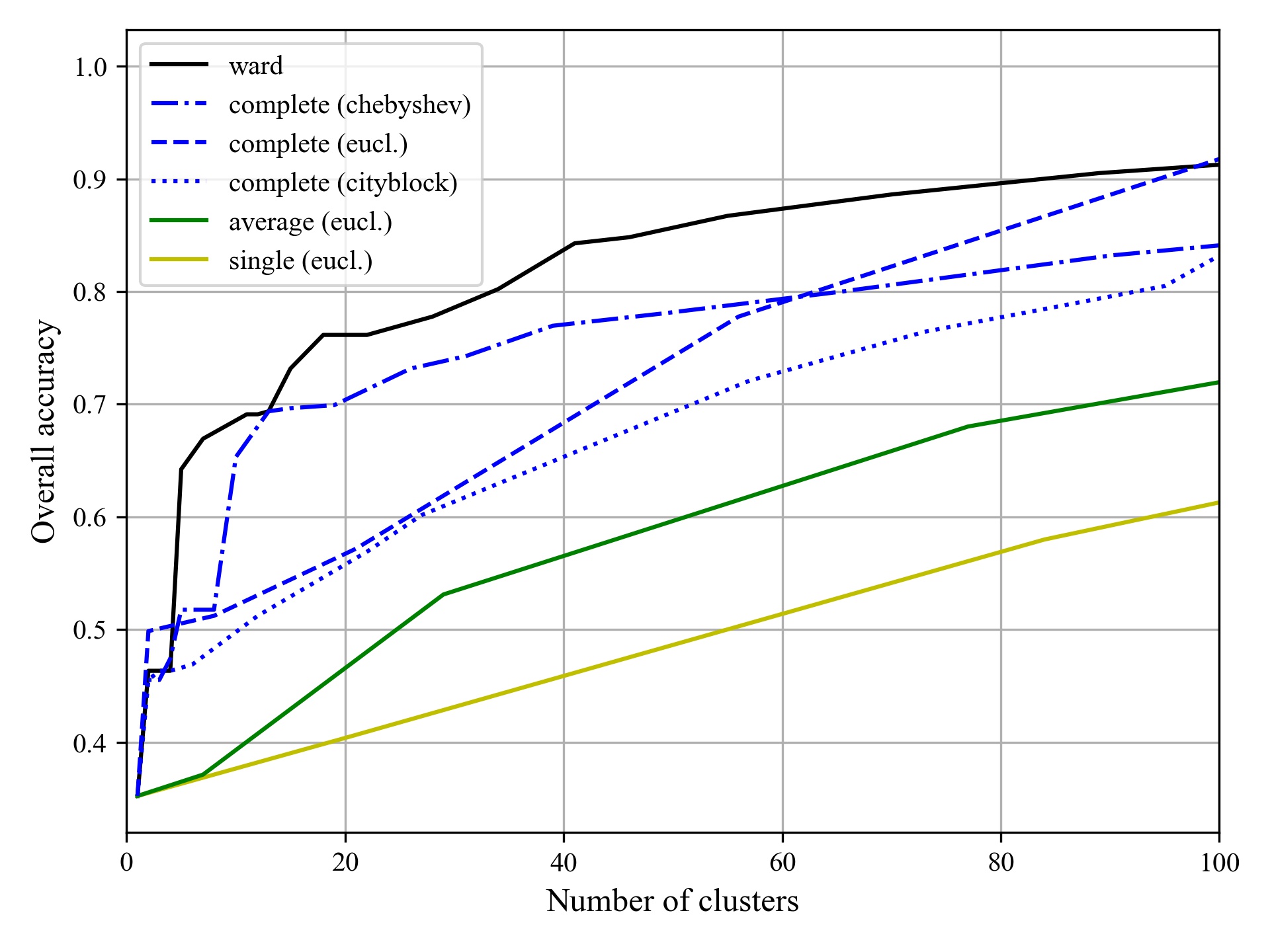}
	\caption{Overall accuracy over the number of clusters for various combinations of linkage criterion and distance metric for hierarchical agglomerative clustering (HAC) at Bendplatz traffic space.}
	\label{fig_linkage}
\end{figure*}

Based on the previous findings, hierarchical agglomerative clustering is used as clustering approach. To determine the most powerful combination of linkage criterion and distance metric for this exemplary application, Figure~\ref{fig_linkage} shows the respective overall accuracy over the number of clusters for each combination within the hierarchical agglomerative clustering approach. Since the focus of this chapter is on the exemplary evaluation of the method, this figure serves for the choice of hierarchical agglomerative clustering in connection with ward-linkage. For possible reasons for the different accuracies of the remaining combinations of linkage criterion and distance metric, please refer to the literature~\cite{Xu.2008,Everitt.2011}.

It should be noted that the labels of the rule-based approach used for determining the accuracy do not represent the ground truth, either. Thus, false positive as well as false negative ratings of the accuracy can occur. Nevertheless, this approach offers a building block for determining a sufficient number of scenario clusters with respect to a reasonable balance between scenario coverage and effort during validation and testing of an ADS.

\subsubsection{Visual Validation}
\begin{figure*}[t]
	\centering
	\includegraphics[width=7in]{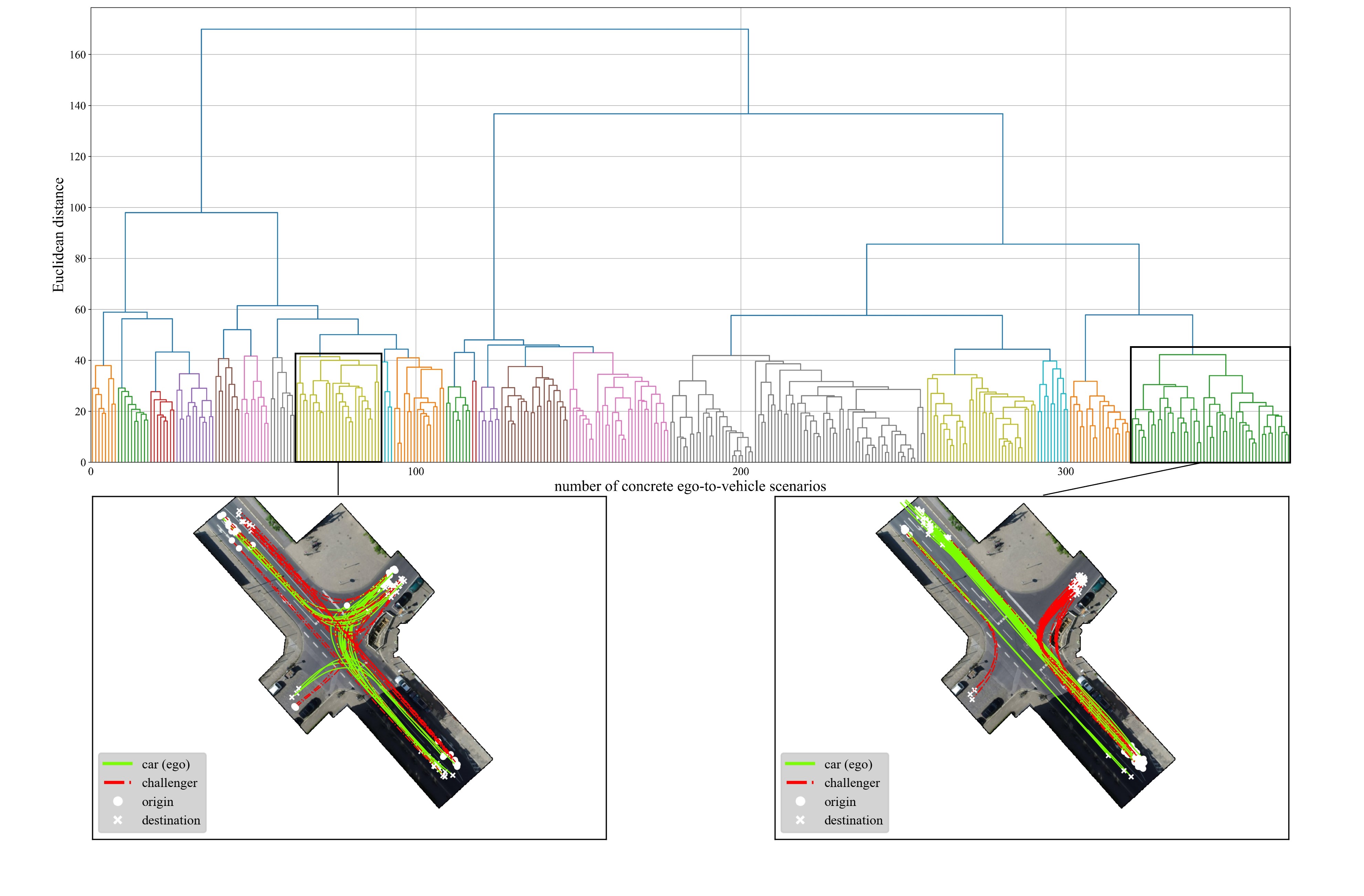}
	\caption{Dendrogram (top) depicting the clustering structure of ego-to-vehicle scenarios at Bendplatz traffic space (color-coded for a partitioning into 20 clusters). The lower part illustrates the mapping of two exemplary clusters to the underlying ego and nearest challenger trajectories in form of cluster validation plots.}
	\label{fig_dendro_v2v}
\end{figure*}

Hierarchical agglomerative clustering follows a bottom-up approach, starting with N clusters, each of which includes exactly one sample. A series of merge operations follows based on a linkage criterion until a threshold is reached or all samples are forced to the same cluster~\cite{Xu.2008,Kerber.2020}. The results of hierarchical agglomerative clustering can be visualized by dendrograms, which are characterized by an inverted tree-like structure~\cite{Patel.2019}. While the root node of a dendrogram represents the entire dataset, each leaf can be regarded as one sample. Therefore, the intermediate nodes describe the degree of similarity between samples, and the height of each branch represents the distance between a pair of samples or clusters, or a sample and a respective cluster~\cite{Xu.2008}.

The upper part of Figure~\ref{fig_dendro_v2v} shows the dendrogram for the e-to-v scenarios at Bendplatz, created based on hierarchical agglomerative clustering with ward-linkage for all available recordings. The dendrogram is color-coded for a partitioning into 20 clusters. On the one hand, this shows that the different process steps of the method are consistent up to this point and that the developed feature representation is compatible with static-data based clustering approaches. On the other hand, however, clustering algorithms can always produce a partitioning given a dataset, regardless of whether or not there exists a particular structure in the data~\cite{Xu.2008}. Thus, the subsequent visual cluster validation shall provide a further gain in knowledge and therefore confidence in the clustering results.


In detail, this approach entails the visualization of all ego and (nearest) challenger trajectories of the respective concrete scenarios assigned to a specific cluster in relation to a georeferenced background image of the traffic space under investigation. This kind of visualization is referred to as cluster validation plot in the context of this work. It should be noted that both the feature representation and the clustering consider the varying number of different road users of the respective concrete scenario (cf. Req. M2). This allows the respective concrete scenario, including all relevant road user trajectories in the vicinity of the ego (i.e., region of interest~$a_{\text{gr}}$), to be used for, e.g., scenario-based testing of an automated driving system. For the sake of clarity, the following cluster validation plots show only the corresponding ego and nearest challenger trajectories.

For visual validation, Figure~\ref{fig_dendro_v2v} illustrates additionally the relationship between individual clusters of the dendrogram and the corresponding cluster validation plots for the e-to-v scenarios at Bendplatz traffic space. To analyze the clustering results, both the yellow cluster on the left side with the smallest accuracy of 0.35 and the green, second-largest cluster on the right side of the dendrogram with an accuracy of 0.98 serve as example for the partitioning into a total of 20 clusters with an overall accuracy of 0.76. Additionally, Table~\ref{tab2} provides an overview of the respective cluster size and accuracy per cluster for the exemplary partitioning. The green cluster (No. 20 according to Table~\ref{tab2}) with nearly optimal accuracy contains 48 branch scenarios and one follow scenario. The reason for the assignment of the follow scenario to this cluster could be due to the exceptional in-lane variability of the vehicles involved in this concrete scenario. However, the existing implementation can extract semantically similar concrete scenarios in which one vehicle follows the lane while the other vehicle turns during the scenario very well and identify them as similar. In addition, this example illustrates the direction independence of the feature representation, as semantically similar scenarios are clustered regardless of their specific position within the traffic space.

Looking at the cluster validation plot of the smaller yellow cluster (No. 8 according to Table~\ref{tab2}) in the left part of Figure~\ref{fig_dendro_v2v}, further conclusions can be drawn about the performance of the current implementation, which appears to depend on the constellation and dynamics of the scenarios under investigation. The distribution of the labels of the rule-based approach for the samples assigned to this cluster is dominated on the one hand by follow scenarios and on the other hand by scenarios in which at least one vehicle involved in the scenario turns (labels 4, 6, and 7). Since exactly these scenario subgroups are assigned almost optimally when the distance threshold is reduced to 39 and the number of clusters is increased to 32, respectively, this emphatically shows the trade-off between cluster accuracy and testing effort. Moreover, the increase in the number of clusters is accompanied by a higher minimum accuracy, which corresponds to a squeeze of the accuracy probability distribution. This is further reflected by the proportion of clusters with the optimal accuracy value of 1, which is 45\% for 20 clusters and 65\% when moving to 32 clusters. More in-depth investigations regarding label-specific performance or possible false positive as well as false negative ratings of the accuracy are subject of ongoing work.

\begin{table}[h]
	\caption{Overview of cluster size and accuracy per cluster for exemplary partitioning into 20 clusters (e-to-v scenarios at Bendplatz traffic space). The numbering of the clusters is done according to the color coding of the dendrogram in Figure~\ref{fig_dendro_v2v} from left to right.}
	\label{tab2}
	\begin{center}
			\begin{tabular}{|l|l|l|}
					\cline{1-3}
					Cluster number & Cluster size & Cluster accuracy \\ \cline{1-3}
					1              & 8                                           & 0.75          \\ \cline{1-3}
					2              & 10                                          & 1                 \\ \cline{1-3}
					3              & 8                                           & 1                  \\ \cline{1-3}
					4              & 12                                          & 1                  \\ \cline{1-3}
					5              & 8                                           & 0.88              \\ \cline{1-3}
					6              & 9                                           & 1                 \\ \cline{1-3}
					7              & 8                                           & 1                 \\ \cline{1-3}
					8              & 26                                          & 0.35              \\ \cline{1-3}
					9              & 4                                           & 1                \\ \cline{1-3}
					10             & 16                                          & 0.44              \\ \cline{1-3}
					11             & 8                                           & 0.75              \\ \cline{1-3}
					12             & 2                                           & 1                \\ \cline{1-3}
					13             & 7                                           & 1                 \\ \cline{1-3}
					14             & 21                                          & 0.48               \\ \cline{1-3}
					15             & 31                                          & 0.74               \\ \cline{1-3}
					16             & 79                                          & 0.76              \\ \cline{1-3}
					17             & 34                                          & 0.56              \\ \cline{1-3}
					18             & 10                                          & 0.7               \\ \cline{1-3}
					19             & 19                                          & 1                 \\ \cline{1-3}
					20             & 49                                          & 0.98              \\ \cline{1-3}
				\end{tabular}
		\end{center}
\end{table}

\begin{figure}[t]
	\centering
	\includegraphics[width=3.2in]{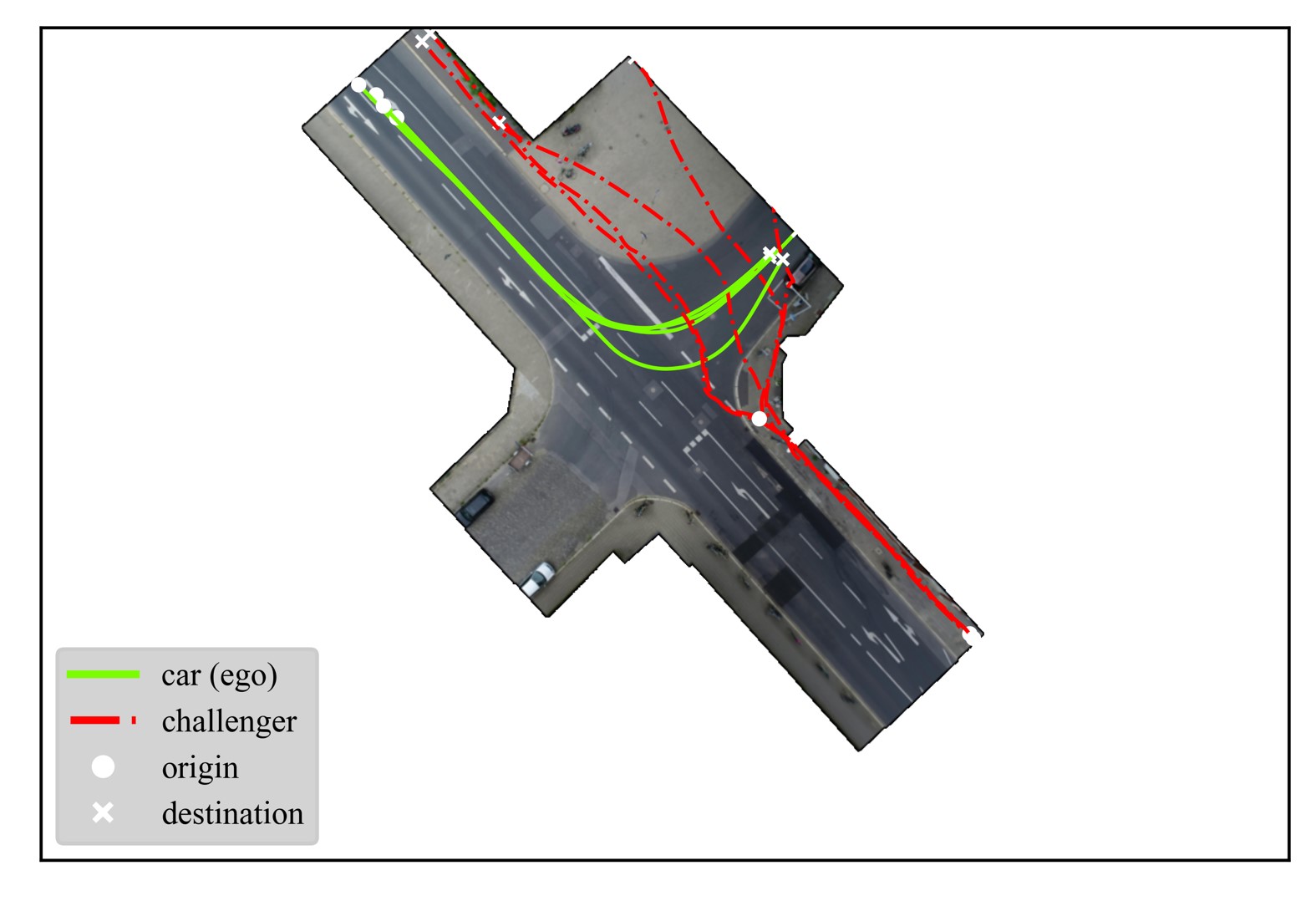}
	\caption{Ego-to-pedestrian cluster validation plot containing the underlying ego and nearest challenger trajectories randomly chosen from one cluster at Bendplatz traffic space.}
	\label{fig_ped}
\end{figure}

While 369 concrete e-to-v scenarios are available for the clustering at Bendplatz traffic space, a significantly lower number of 33 e-to-p scenarios remains as clustering input after spatiotemporal filtering. In analogy to the procedure for the e-to-v scenarios, Figure~\ref{fig_ped} shows a cluster validation plot for e-to-p scenarios at Bendplatz. The preliminary assessment that the extraction method is capable of clustering semantically similar scenarios is strengthened. This assessment also applies for the case of multimodal interaction in Figure~\ref{fig_ped}, including less rule-based pedestrian behavior.


\section{Discussion of the Requirements Set for the Scenario Extraction Method} \label{discussion}

We developed the method for extracting multimodal urban traffic scenarios relevant for testing automated driving systems based on the requirements stated in Section~\ref{Background}. The present section serves the purpose of evaluating the fulfillment of these requirements. The majority of the requirements that have been classified as must depends significantly on the implementation of the feature representation. The proposed scenario tensor comprises a discrete, multi-channel grid structure, representing the spatiotemporal scenario characteristics from a bird´s-eye view. This feature representation enables the processing of naturalistic road traffic data by means of (multivariate) time series of different lengths and includes multiple road user dynamics simultaneously, so we consider requirements M1 and S1 as fulfilled.
 
Furthermore, the search algorithm within the step of spatiotemporal filtering is capable of slicing the time series data of a corresponding ego-vehicle into several sub-sequences or concrete scenarios, respectively, which leads to the fulfillment of requirement M5. Since the exemplary evaluation in Section~\ref{Eval} showed that the method allows the clustering of multimodal urban traffic scenarios of a varying number of road users, we also regard requirement M2 as fulfilled. Further, we also consider requirement M3 as fulfilled, which demands the applicability of the method to different urban traffic areas with as little adaptation effort as possible. This is mainly shown by exemplarily applying the method to the inD and the sv dataset, which in total comprise eleven traffic spaces. While the results of the principal feature selection were presented explicitly for all traffic spaces, Bendplatz traffic space was used for an in-depth analysis of the subsequent steps, where a higher proportion of multimodal interaction can be observed. The corresponding results have shown that the method allows for clustering semantically similar scenarios without including road network information, which leads to the conclusion that requirement C1 is fulfilled. It should be noted that georeferenced background images of the traffic spaces under investigation are required to perform the visual validation of the clustering results using the cluster validation plots.

Since the developed feature representation is technically not significantly different from established approaches in the field of computer vision, requirement M4 may also be considered fulfilled. As far as a dataset is available that contains the trajectories of all road users with sufficient accuracy for the traffic space to be examined, we are not aware of any structural limitation of the method with respect to the application to other traffic spaces (regardless of the technical implementation for generating the corresponding dataset). The requirement for the availability of the trajectories of all road users in the dataset is equally related to the fulfillment of requirement W1.

In addition, it is conceivable that the cluster results of the method are sensitive with respect to different weather conditions (layer five of PEGASUS six-layer model) if the previously described information is available in the given dataset. This is the case because significantly different weather conditions in the dataset will also result in a different feature representation. Specifically, changing weather conditions lead to changes in road user behavior, such as lower deceleration or cornering speeds during heavy rainfall. This in turn leads to changed velocity grids and acceleration grids, which is reflected in a different clustering structure. It is conceivable that in case of a low distance threshold value, concrete scenarios of a functional scenario type with similar weather conditions form a cluster, initially. When increasing the distance threshold value, these concrete scenarios form a cluster with concrete scenarios of the same functional scenario type but deviating weather conditions (assuming that these scenarios are present in the dataset). Finally, with further increase of the distance threshold value, different functional scenarios form a cluster. It should be noted that the verification of this statement by experiments is yet to be done. The same applies to the answer to the question how much expert knowledge is necessary to apply the method (cf. Req. S2).

\section{Conclusion and Outlook} \label{conclusion}


This paper proposes a method for extracting multimodal urban traffic scenarios from naturalistic road traffic data in an unsupervised manner relevant for the safety validation of automated driving systems. The method comprises five process steps, namely spatiotemporal filtering, principal feature analysis, feature extraction, clustering, and cluster validation. The results of the principal feature analysis show a dependence of the cumulative explained variance within the data on both traffic space and road user type. In addition, the required features for the subsequent steps can be reduced by a factor of two within the exemplary evaluation. A discrete, multi-channel grid structure for scenario modeling is proposed, resulting in a scenario tensor containing various grid matrices defined by the previously selected features. Based on the developed feature representation, it is possible to apply clustering methods relying on static data. To evaluate the method, the performance of different clustering approaches is compared, before discussing results of hierarchical agglomerative clustering applied to an urban traffic space. Thereby, the results of both the comparison of different clustering approaches with respect to a rule-based baseline approach and the visual validation for selected multimodal urban traffic scenarios utilizing so-called cluster validation plots are promising depending on the number of clusters. Moreover, the observed jump in overall accuracy of the hierarchical agglomerative clustering of around 20\% when moving from 4 to 5 clusters and the saturation effect starting at 41 clusters with an overall accuracy of 84\% can be a valuable contribution in the context of the trade-off between the number of functional scenarios (i.e., clustering accuracy) and testing effort. In detail, the clusters, whose number depends on the required overall accuracy, can be used to derive equivalence classes or functional scenarios of a test catalog. Thus, the method can be a valuable support for the scenario selection argumentation within comprehensive validation strategies in the context of relevant standards or regulations for automated driving systems.

Based on the results, future research should push for a broader evaluation of the method by investigating more traffic spaces. Optimally, datasets examined in the future will contain explicit information about features beyond the object layer (layer 4 of PEGASUS six-layer model~\cite{Scholtes.2021}) to be able to prove the sensitivity of the scenario extraction method, e.g., with respect to different weather conditions (layer 5 of PEGASUS six-layer model). Furthermore, in-depth studies are needed to further increase the clustering accuracy for each defined cluster number. Potential entry points could be the adaption of the position of the ego-vehicle in the respective scenario grid or a nonlinear feature extraction, e.g., by convolutional autoencoders. Moreover, the incorporation of the implemented method into the scenario-based simulation platform presented in~\cite{Weber.2021} represents a future use case. Thereby, the scenarios extracted by the method represent one main input channel for the adaptive replay-to-sim approach (ARTS), in addition to an agent-based simulation and the automated driving system under test. Finally, the method will be integrated into a semi-supervised machine learning pipeline to achieve the medium-term goal of a robust and generic scenario classifier. The presented method contributes to the quantitative real-data based extraction of relevant traffic scenarios. This method can be seen as building block toward a systematic, data-driven construction of a relevant scenario database with sufficient coverage in a fully automated manner for validating the safe behavior of automated driving systems.

\addtolength{\textheight}{-0cm}





\section*{ACKNOWLEDGMENTS}

We want to thank Ulrich Eberle and Stefan Berger (Opel Automobile GmbH, Stellantis NV) for the productive discussions and the feedback on our approaches as well as for peer review of this publication prior to its submission.


\bibliographystyle{IEEEtran}%
\bibliography{IEEEabrv,literature}%

 
\end{document}